# Genetic Algorithm with a Local Search Strategy for Discovering Communities in Complex Networks

**Dayou Liu**[1, 4], **Di Jin**[2*], **Carlos Baquero**[3], **Dongxiao He**[1, 4], **Bo Yang**[1, 4], **Qiangyuan Yu**[1, 4]

[1] *College of Computer Science and Technology, Jilin University, Changchun, 130012, China*
[2] *College of Computer Science and Technology, Tianjin University, Tianjin, 300072, China*
[3] *HASLab, INESC TEC & University of Minho, Braga, Portugal*
[4] *Key Laboratory of Symbolic Computation and Knowledge Engineering of Ministry of Education, Jilin University, Changchun 130012, China*
E-mail: jindi@tju.edu.cn

**Abstract**

In order to further improve the performance of current genetic algorithms aiming at discovering communities, a local search based genetic algorithm (GALS) is here proposed. The core of GALS is a local search based mutation technique. In order to overcome the drawbacks of traditional mutation methods, the paper develops the concept of marginal gene and then the local monotonicity of modularity function $Q$ is deduced from each node's local view. Based on these two elements, a new mutation method combined with a local search strategy is presented. GALS has been evaluated on both synthetic benchmarks and several real networks, and compared with some presently competing algorithms. Experimental results show that GALS is highly effective and efficient for discovering community structure.

*Keywords*: Complex network; Community mining; Network clustering; Genetic algorithm; Local search; Modularity $Q$

## 1. Introduction

Many complex systems in the real world exist in the form of networks, such as social networks, biological networks, Web networks, etc, which are collectively referred to as complex networks. The area of complex networks has attracted many researchers from different fields such as physics, mathematics, computer science, etc. While a considerable body of work addressed basic statistical properties of complex networks, such as the existence of a "small world effect" [1] and the presence of "power laws" in the link distribution [2], another property has also been given particular attention, that is, "community structure": where the nodes in networks are often found to cluster into tightly-knit groups with a high density of within-group edges and a lower density of between-group edges [3]. The community mining problem (CMP), which this paper refers to, is to discover and interpret community structures from various complex network data.

The ability to detect community structure is useful in many aspects [4]. For example, nodes belonging to the same community may have much more common features than those in different communities, which could be used to simplify the functional analysis of complex networks. Furthermore, community structure may provide insights in understanding some uncharacteristic property of a complex network system. For instance, in the world wide web, community analysis has uncovered thematic clusters; in biochemical or neural networks, communities may be functional groups and separating the network into such groups could simplify functional analysis considerably.

In the past few years, the most popular method to detect communities in graphs consists in the optimization of a quality function, modularity $Q$ introduced by Newman and Girvan [5]. Modularity $Q$ gives a clear and precise definition of the characteristics of the acknowledged community and has had very successful application in practice [6], although it is still not free from drawbacks (suffers from resolution limits [7], exhibits extreme degeneracy [8], has random graphs with high modularity due to fluctuations [9], etc).

---

[*] Corresponding author.

The search for the partition with maximal modularity is in general a great challenge since it was proved to be a NP-hard (non-deterministic polynomial-time hard) problem [10]. Many heuristics relying on different approaches have been introduced to approximate the optimal partition, and some of them are able to find fairly good approximate solutions in a reasonable time. But there is still room for improvement in their performance, in terms of both effectiveness and efficiency.

Because of the effectiveness in approaching NP-hard problems, genetic algorithms have become a competitive community mining method. In order to further improve the performance of those former genetic algorithms for CMP, a local search based genetic algorithm GALS is proposed in this paper. Our GALS employs modularity $Q$ as objective function, and takes a graph-based representation LAR [11] as genetic representation. In GALS, we first adopt a Markov random walk based method to generate the initial population; and then we detect community structure by iteratively executing the following three genetic operators: uniform crossover [12], local search based mutation and $\mu+\lambda$ selection [13]. Moreover, the genetic operators, which are employed by GALS, make each LAR chromosome in the population correspond to a spanning subgraph of the original network. Thus the solution space of GALS can be reduced, which improves both the search efficiency and convergence rate of this algorithm.

It is noteworthy that, the core of GALS is the local search based mutation method. For overcoming the drawbacks of traditional mutation methods, we first present and name the concept of marginal gene, and then deduce the local monotonicity of modularity $Q$ from each node's local view. Based on these above two points, an effective and efficient mutation method combined with a local search strategy is finally proposed.

## 2. Related Works

Over the last decade, many approaches have been proposed to the analysis of the community structures in complex networks. They adopt different types of principles and techniques, rooted in physics, mathematics, computer science, and so on. They mainly include: **divisive methods**, e.g. Girvan-Newman (GN) algorithm [3]; **modularity optimization methods**, e.g. Fast Newman (FN) algorithm [14], Simulated Annealing (SA) algorithm [15], Fast Unfolding Algorithm (FUA) [16]; **label passing methods**, e.g. Label Propagation Algorithm (LPA) [17], hub-based algorithms [18, 19]; **dynamic methods**, e.g. Finding and Extracting Communities (FEC) [20], Infomap algorithm [21], Ronhovde and Nussinov (RN) algorithm [22]; and others. The interested readers can consult the excellent and comprehensive survey by Fortunato [6].

Especially, because of the effectiveness to approximately solve NP-hard problems, genetic algorithm (GA) is currently becoming a class among the competitive methods for modularity optimization. At present two main types of genetic representation strategy are employed by GA in solving community mining problems.

The first one is a string-of-group encoding (SGE). As each node is only denoted by an arbitrary label in a SGE chromosome, the traditional crossover operators (such as uniform, one-point and two-point crossover) is not well-fit for the task. This placed pressure on the design of new types of crossover operators. Several works in this line are as follows. Tasgin et al. [23] presented, for the first time, the use of a genetic algorithm to detect communities. In their method, a one-way crossover operator is proposed, which has been proven to be effective for this string-of-group encoding. He et al. [24] proposed a genetic algorithm with ensemble learning for discovering communities. They replaced the traditional crossover operator with a new multi-individual crossover operator based on the idea of clustering ensemble, which also performs very well. Li et al. [25] proposed a genetic algorithm for community detection, which adopts the one-way crossover operator and introduces a new type of local search operator. However, the above methods are all only effective for some small sized benchmarks. Thus, the research community began to focus on a second type of encoding strategy.

The second one is called the locus-based adjacency representation (LAR), which is indeed a graph-based representation. As each community corresponds to a component in any a LAR chromosome, this type of representation strategy is well-fit for most of the traditional crossover operators. There are some related works as follows. Pizzuti et al. [26] proposed a genetic algorithm using LAR representation for the first time, which adopts the traditional operators (such as uniform crossover and random mutation) for the detection of communities. Later, they further improved their former method by employing two types of community functions, and presented a multi-objective genetic algorithm [27]. But the performance of both these two approaches also looks only effective under the small benchmarks as before. Shi et al. [28] proposed another genetic algorithm with LAR representation, in which a new crossover operator, similar with the one-way crossover operator, is introduced. As far as we know, this is the first genetic algorithm that is able to cluster several large networks. However, its clustering quality is still not ideal when compared with some current competing algorithms for community detection.

In our recent study, we found out that, the traditional random mutation operator often results in merging and/or splitting communities for the LAR representation, although it has some advances for the crossover operation. This will make it unable to effectively achieve its local search function, and then lead to the inefficacy of genetic algorithms. Thus, in our opinion, designing an efficient and effective mutation operator for the LAR representation may be a good solution, which may make a genetic algorithm have stronger ability to deal with actual large-scale networks.

## 3. Algorithm

### 3.1. *Problem definition*

Let $N = (V, E)$ denote an unweighted and undirected network, where $V$ is the set of nodes (or vertices) and $E$ is the set of edges (or links). Let a $k$-way partition of the network be defined as $\pi = \{c_1, c_2, \ldots, c_k\}$, where $c_1, c_2, \ldots, c_k$ denote the $k$ clusters, and satisfy $\bigcup_{1 \leq i \leq k} c_i = V$ and $\bigcap_{1 \leq i \leq k} c_i = \varnothing$. If partition $\pi$ has the property that within-cluster edges are dense and between-cluster edges are sparse, it's called a community structure of this network. This is a way of defining non-overlapping communities. However, it should be noted that, the communities may overlap in many real-world networks, i.e. some nodes may belong to more than one community simultaneously [29]. Here we mainly focused on hard clustering, and did not consider overlapping communities in this present work.

In 2004, Newman and Girvan proposed an important quality metric for assessment of partitioning a network into communities, which is called modularity $Q$ [5]. The idea of modularity $Q$ steams from the intuition that a network with community structure is different from a random network. Therefore, this function $Q$ can be defined as the difference between the fraction of edges that fall within communities and the expected value of the same edge quantity if edges fell at random without regard for the community structure. As modularity $Q$ has been widely accepted by the scientific community [6], in this paper we also choose to employ it as the objective function that is to be maximized.

Given the network $N$ and supposing its nodes are divided into some communities such that node $i$ belongs to community $c_{r(i)}$ in which $r(i)$ denotes the label of node $i$, then function $Q$ is defined as

$$Q = \frac{1}{2m} \sum_{ij} \left( \left( A_{ij} - \frac{k_i k_j}{2m} \right) \times \delta(r(i), r(j)) \right). \quad (1)$$

Here $A = (A_{ij})_{n \times n}$ denotes the adjacency matrix of network $N$, in which $A_{ij} = 1$ if nodes $i$ and $j$ connect with each other, $A_{ij} = 0$ otherwise. The $\delta$ function $\delta(u,v)$ is equal to 1 if $u = v$ and 0 otherwise. The degree $k_i$ of any node $i$ is defined as $k_i = \sum_j A_{ij}$, and the total number of edges $m$ in this network is defined as $m = \frac{1}{2} \sum_{ij} A_{ij}$.

### 3.2. *Genetic representation*

Algorithm GALS in this paper adopts the locus-based adjacency representation (LAR) proposed by [11]. At present, this representation (or encoding) schema was also employed by [30, 31] for multi-objective clustering problem and by [26-28] for community mining problem. In this graph-based representation, any individual (chromosome) $g$ in the population consists of $n$ genes, in which each gene corresponds to a node in network $N$ and $n$ denotes the total number of nodes in this network. Each gene $i$ can take an arbitrary allele value $j$ in the range of $\{1, …, n\}$, which can be interpreted as a link between nodes $i$ and $j$ existing in the corresponding graph $G$ of individual $g$. This also means that, node $i$ will be in a same community with node $j$ in the partition denoted by this individual. The decoding process for a LAR individual is to identify all the components from graph $G$, and the nodes belonging to the same component are assigned to a same community. This decoding process can be done in a linear time as shown by [32]. A simple example of the LAR representation is illustrated as Fig. 1.

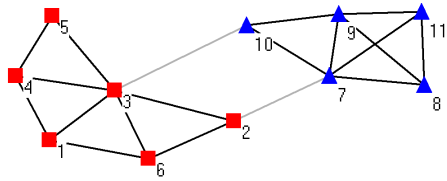

(a)

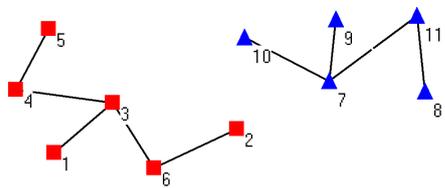

(b)

(c)

**Fig. 1.** (Color online) An illustration for the locus-based adjacency representation. (a) A sample network consisting of eleven nodes; (b) One out of many possible chromosomes; (c) The corresponding graph of this chromosome.

The locus-based adjacency representation has two major advantages for solving community mining problems. Firstly, the community number denoted by each individual is automatically determined in the decoding process, thus there is no need for us to know the number of communities in advance. Hence, we can effortlessly evaluate a network clustering solutions with different community numbers during the execution process of GALS, and finally attain the best solution with the most suitable community number. Secondly, LAR representation is well suited for standard crossover operators, such as uniform, one-point and two-point crossover. For community mining problems, if we use the traditional genetic representations such as string-of-group encoding, the above crossover operators will be highly disruptive, as well as detrimental, for the community detection process. In contrast, the graph-based representation can effortlessly implement its global search function by merging and splitting communities, while maintaining good heritability.

### 3.3. *Population initialization*

Here we first introduce the idea of safe individuals proposed by [26], and then give a Markov random walk based individual generation method which can produce safe, accurate and diverse initial individuals.

If an individual is randomly generated, some components in its corresponding graph $G$ may be disconnected in the original network $N$, which also means $G$ may be not a subgraph of $N$. For example, an individual could contain an allele value $j$ in the $i$th position, which means there is a link between nodes $i$ and $j$ in its corresponding graph $G$, but there may be no connection between these two nodes in network $N$. However, in a network with community structure, there is the obvious intuition that any a node should have one of its neighbors in the same community or it, itself, is a community. Thus, the solution space of chromosomes can be reduced if the above heuristics is considered, which means one can make any individual in the population be a spanning subgraph of the original network $N$. Then the individual that satisfies the above condition is called a safe individual, the population that is composed of safe individuals is called a safe population, and the solution space of safe individuals is called a safe solution space.

The whole solution space of LAR representation is $n^n$, while the safe solution space is $\prod_{i=1}^{n} k_i$, where $n$ is the total number of nodes and $k_i$ is the degree of node $i$ in network $N$. As most complex networks are sparse graphs, for any node $i$, $k_i$ can be regarded as a constant ($k_i \ll n$). Thus it is obvious that the safe solution space is much smaller than the whole solution space. Therefore, if the search region of our algorithm GALS can be restricted to the range of the safe solution space, its search efficiency and convergence rate can both be improved.

In order to achieve the above goal, we propose a Markov random walk based individual generation method (MRW), which is based on the natural community property possessed by complex networks.

In a network, let $q_{ij}$ be the probability that an agent freely walks from an arbitrary node $i$ to one of its neighbor nodes $j$ within one step, which is also called the transition probability of one-step random walk. In terms of the adjacency matrix of $N$, $A = (A_{ij})_{n \times n}$, $q_{ij}$ is defined by (2). Here the agent's random walk can be regarded as a discrete Markov process.

$$q_{ij} = \frac{A_{ij}}{\sum_r A_{ir}}. \qquad (2)$$

From the view of a Markov random walk, when a network has community structure, a random walk agent should find it difficult to move outside its own community boundary, whereas it should be easy for it to reach other nodes within its community, as link density within a community should be high, by definition. In other words, the probability for remaining in the same community, that is, a random walk agent starts from any node and stays in its own community, should be greater than that of going out to a different community. The readers, who are interested in heuristic based random walks to find communities, can consult the related and excellent work by Pons and Latapy [33].

Based on the above idea, in algorithm MRW, we make any gene $i$ in a chromosome select its allele value $j$ in the range of $\{1, \ldots, n\}$ by using the one-step transition probability $q_{ij}$. It's obvious that the individuals generated by MRW are not only safe but also accurate and diverse, which can improve the performance of our genetic algorithm GALS.

### 3.4. *Selection and crossover operators*

The role of a selection operator is to sieve out (deterministically or probabilistically) the solution that have been generated through the use of the other genetic operators. In order to keep the fittest individuals from each generation and improve the convergence speed of genetic algorithm (GA) at the same time, a $\mu+\lambda$ selection strategy [12], preferred by GA for solving combinatorial optimization problems, is employed in this paper. The process of $\mu+\lambda$ selection can be described as follows. Let the size of parent population be $\mu$, and $\lambda$ offspring be generated from randomly chosen parents, then we single out $\mu$ best individuals among parents and offspring as the population of next generation.

As a global search operator in GA, uniform crossover (UC) [13] is adopted in this paper. Given two randomly chosen parents $A$ and $B$, and a randomly generated binary vector $v$, uniform crossover then selects the genes where $v$ is 1 from parent $A$, and selects the genes where $v$ is 0 from parent $B$, and then combines the genes to form a new child $C$. Mathematically speaking, one has $C = A.*v + B.*(1-v)$, where $(.*)$ denotes array multiplication. It's obvious that if parents $A$ and $B$ both are safe individuals, their child $C$ is forcibly a safe one too. This is because any gene $i$ containing a value $j$ in the child must comes from one of its two parents, meanwhile, this edge $<i, j>$ will necessarily exist in network $N$ due to the safety of the parents. Thus, we find out that the uniform crossover will not violate the safety of the population in GA. Furthermore; here we choose the uniform crossover in favor of one-point or two-point crossover. This is because it is unbiased with respect to the ordering of genes and can generate any combination of alleles from the two parents within a single crossover operation. An example for the operation of uniform crossover on LAR chromosomes is shown as Fig. 2.

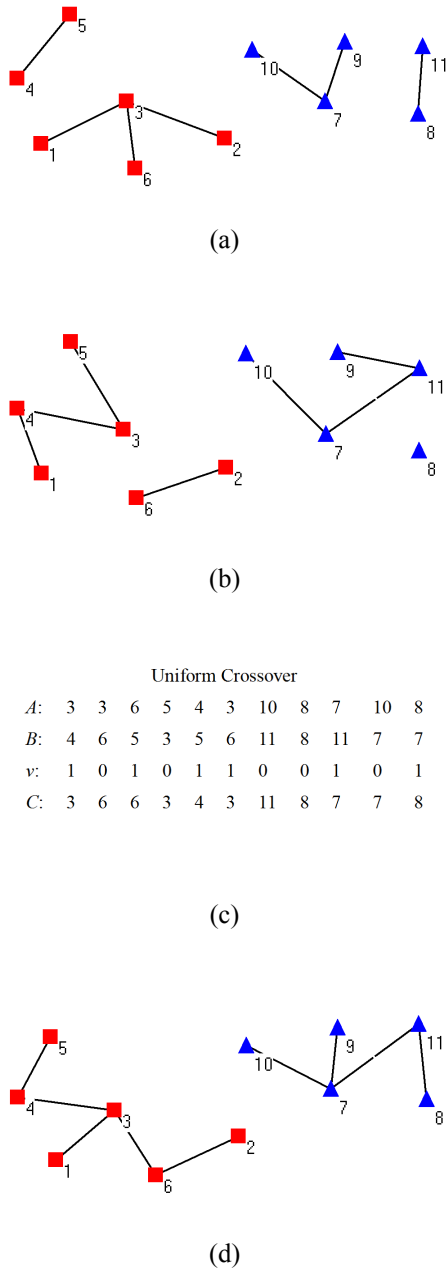

**Fig. 2.** (Color online) An illustration of the uniform crossover on the network in Fig. 1(a). (a) The corresponding graph of parent *A*; (b) The corresponding graph of parent *B*; (c) Uniform crossover of parents *A* and *B* yielding child *C* by using random binary vector *v*; (d) The corresponding graph of child *C*.

### 3.5. *Mutation operator*

The mutation operator is the most important part in this paper. Here, for the community mining problem, we first introduce a concept called *marginal gene* intended to overcome the drawbacks of current mutation methods. Then, we propose an efficient and effective local search based mutation algorithm, which operates with the aid of this marginal gene.

#### 3.5.1. *Marginal gene*

Recently, the random mutation strategy was mainly adopted as a local search operator by most genetic algorithms which employed LAR representation [26-28, 30, 31]. However, this type of mutation operator is not well suited for the community mining problem.

In the opinion of Guimera and Amaral, when solving community mining problems, it's an effective method to generate a new candidate solution by iteratively executing the following three types of operations on the current candidate solution, which includes "move single nodes from one community to another", "merge multi-communities" and "split single communities" [15]. In a genetic algorithm, the crossover operator is regarded as a macroscopic operation on individuals, while the mutation operator is regarded as a microscopic operation on individuals. Thus, in a genetic algorithm for solving community mining problems, if the crossover operator can achieve its global search function by merging and splitting communities, and the mutation operator can achieve its local search function by moving single nodes between communities, this genetic algorithm leads to a strong ability in searching. Each individual in this paper corresponds to a graph, and each component in the graph corresponds to a community. Thus it's obvious that the uniform crossover in Sec. 3.4 can effortlessly achieve its function of merging and splitting communities. Then, the mutation operator in this section should be requested to have the ability to effectively achieve its intended function of moving single nodes between communities.

However, our study shows that the traditional mutation operator [26-28, 30, 31] often results in merging and/or splitting communities, which will make it unable to effectively achieve its local search function, and then lead to the inefficacy of a genetic algorithm. For example, in an individual (or chromosome) *g*, let nodes *i* and *j* belong to different communities (or components), and gene *i* select the *j* as its allele value by a single mutation operation, then this mutation operation will very likely lead to the combination of these two communities, which also means the two components denoted by nodes *i* and *j* may be merged into a bigger one by building a new link <*i*, *j*> between them. A simple example is shown as Fig. 3(c). It's obvious that this is an undesirable situation. Thus, as the following presentation will show, we will offer an effective solution to this problem.

**Definition 1. (Marginal Gene)** *Given an arbitrary LAR chromosome g, if the allele values of all genes in g are not equal to j, gene j is called a marginal gene in g, which is also called a marginal node.*

Note that any LAR chromosome *g* corresponds to a *directed* graph *G*, although we depict it as an undirected graph in the above sections for simplification. All nodes in *G* will not point at marginal nodes known from Definition 1. Thus, a single mutation operation on a marginal node can implement this node's movement from one community to another. Meanwhile, it will not result in merging or splitting communities. It's obvious that, with the aid of marginal nodes, the mutation operator can successfully achieve its local search function. A simple example which describes a single mutation operation on an arbitrary marginal node is shown as Fig. 3(d).

| Position | 1 | 2 | 3 | 4 | 5 | 6 | 7 | 8 | 9 | 10 | 11 |
|---|---|---|---|---|---|---|---|---|---|---|---|
| Genotype | 3 | 6 | 6 | 3 | 4 | 3 | 11 | 8 | 7 | 7 | 8 |

(a)

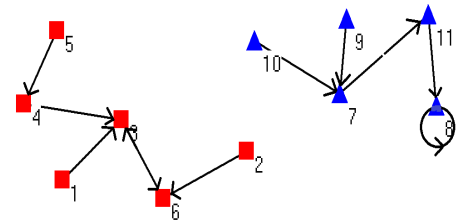

(b)

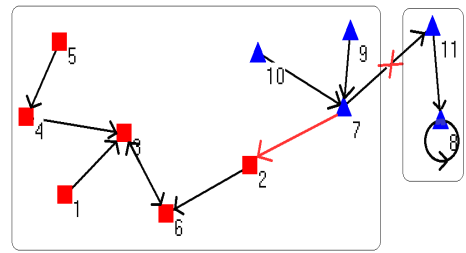

(c)

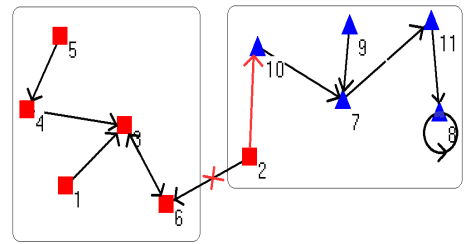

(d)

**Fig. 3.** (Color online) An illustration of the mutation operation for the network in Fig. 1(a). Note that, in order to depict the mutation process more clearly, here we adopt the *directed* graph to express a chromosome. (a) An arbitrary chromosome *A*; (b) Its corresponding *directed* graph *G*; (c) A single mutation operation on non-marginal node 7 for chromosome *A*, which results in splitting and/or merging communities; (d) A single mutation operation on marginal node 2 for chromosome *A*, which implements this node's movement from the red community to the blue one, and does not result in merging or splitting communities.

Now, we have to pay attention to the proportion of marginal genes that appear in a LAR chromosome. We first focus on the randomly generated chromosome *g*, and then generalize it to a more universal situation. Let there be *n* genes in *g*, then the probability that any gene *i* takes some value *j* is $p = 1/n$; on the contrary, the probability that gene *i* doesn't take value *j* is $1-p$; and then the probability that all genes in g don't take value j is $\beta = (1-1/n)^n$. Thus the probability that the marginal genes appear in chromosome g should be $\beta$. $\beta(n)$ is a monotonically increasing function, and $\lim_{n\to+\infty} \beta(n) = 1/e \approx 0.3679$. Also, the total number of nodes in almost all real networks is greater than 10, and one has $\beta(10) = 0.3487$. Thus, the proportion of marginal genes appearing in a chromosome g should be $\beta(n) \in (0.3487\sim 0.3679)$, $n > 10$. Moreover, our experiments show that the above conclusion is also fit for more general chromosomes in a universal situation. In this sense, if we execute a mutation operation on all marginal genes for a chromosome, it's equivalent to that of executing a mutation operation on this chromosome by mutation rate $\beta$.

### 3.5.2. Local search based mutation algorithm

In this section, we first deduce the local monotonicity of modularity function Q, and then propose a fast, as well as effective, local search based mutation algorithm (LSMA), constructed with the aid of marginal nodes, as by Definition 1.

In order to cleverly implement single node movements between communities in algorithm LSMA, we now offer some theoretical analyses on our objective function Q, from each node's local view. We convert (1) to (3), where $c_{r(i)}$ denotes the community of node i. It's obvious that, equation (3) makes function Q as the sum of local function f of all nodes. From each node's local point of view, function f can be regarded as the difference between the number of edges that fall within communities and the expected number of edges that fall within communities. Therefore, the function f of each node can measure whether a network division indicates a strong community structure from each node's local point of view. Some Propositions and Theorems on function f are given as follows.

$$Q = \frac{1}{2m}\sum_i f_i, \quad f_i = \sum_{j \in c_{r(i)}} \left( A_{ij} - \frac{k_i k_j}{2m} \right) \quad (3)$$

**Proposition 1.** *For $\forall i \in V$, the local function $f_i$ of any node i in a complex network is only related to its own community $c_{r(i)}$.*

Proposition 1 is obvious according to (3).

**Theorem 1.** *For $\forall i \in V$, if the label of node i changes under the condition that the labels of all other nodes don't change, function Q of the entire network is monotonically increasing with function $f_i$.*

Please see the Appendix part for the Proof of Theorem 1.

Again, there is the simple intuition that, any node in a complex network which has a community structure should have the same label as one of its neighbors, or be in itself is a community. Thus, in our mutation algorithm, any marginal node i can take a node j in the range of $NS_i$ (instead of V) as its allele value, where $NS_i$ denotes the neighbour set of node i and V is the node set of the network.

Further, known from Theorem 1, function Q of a network is monotonically increasing with any node's local function f. In order to further improve the performance of our mutation method, we select here node j from $NS_i$, which can maximize function $f_i$, as the allele value of marginal node i. Here we only consider marginal nodes. Thus, when a single mutation operation is executed on any a marginal node i, the labels of all other nodes will not change, which will necessarily meet the condition which is required by Theorem 1.

Moreover, note that any safe individual is still a safe one after executing our mutation method.

Based on the above discussions, we now present a fast and effective local search based mutation algorithm LSMA, described as Fig. 4.

```
Procedure  LSMA
Global: g /* a chromosome to be mutated */
Begin
1     C ←decode chromosome g // C is the community structure corresponding to g
2     For i=1: n    // n is the total number of genes in g
3         If node i is a marginal node
4             NS_i ←attain all the neighbors of node i // note that NS_i contains node i
5             labels ← get unique labels from NS_i
6             max ← – ∞
7             For each r∈ labels
8                 f_i ←compute function f_i when node i takes label r
9                 If  f_i > max
10                    max ← f_i
11                    label_i ← r
12            End
13        End
14        g(i) ← randomly select a node from NS_i whose label is label_i // update chromosome g
15        C(i)← label_i // update community structure C
16    End
17 End
End
```

**Fig. 4.** LSMA algorithm.

As we can see, in algorithm LSMA, the variation of any a node's allele value denotes the node's movement from its original community (or component) to another one in the corresponding graph of chromosome *g*. Meanwhile, it will not result in merging or splitting communities, as these nodes are all marginal nodes. On the contrary, a mutation operation on non-marginal nodes will necessarily lead to merging or splitting communities for *g*, thus these non-marginal nodes are not suitable for our algorithm LSMA. As we can see, only marginal nodes can meet the requests of LSMA, and this method is especially designed for the marginal nodes.

In order to further explain the effectiveness as well as efficiency of LSMA, some Propositions are given as follows.

**Proposition 2.** *For any chromosome g that adopts LAR representation, the value of fitness function Q will not decrease after executing LSMA algorithm.*

**Proof.** Known from the algorithm flow of LSMA, any marginal gene *i* in chromosome *g* will necessarily take one of its neighbours *j* which can maximize its function $f_i$ as its allele value after a LSMA operation. This means node *i* will move to the community of node *j* and it will not result in merging or splitting communities, which also means node *i* will take the label of node *j* as its new label under the condition that the labels of all other nodes don't change. Known from Theorem 1, if the variation of the label of one node makes its function *f* increase under the condition that the labels of all other nodes don't change, this variation will cause an increase of the *Q*-value of the entire network. Thus, the mutation operation on any marginal gene in chromosome *g* will not make the *Q*-value of the network decrease. □

In network *N*, let the total number of nodes be *n*, the average degree of all the nodes be *k*, and the average community size in the network clustering solution denoted by chromosome *g* in algorithm LSMA be *c*. The time complexity analysis of LSMA is given as follows.

**Proposition 3.** *The time complexity of algorithm LSMA is O(cn).*

**Proof.** It's obvious that the time complexity of the 8th step is the highest in LSMA. Step 8 computes function $f_i$ of each marginal node *i* for all the labels of its neighbours. Because the labels of any node's neighbours are likely to overlap, the average number, to evaluate function *f* for each node *i* in step 8, can't be greater than *k*. Known from Proposition 1, the average time that function *f* is computed once can't be greater than *c*. The total number of marginal nodes is about *βn* in a LAR chromosome. Thus, the time complexity of LSMA can't be greater than *O(βnkc)*. Furthermore, as complex networks are always sparse graphs, which means *k* is a constant, and parameter *β* is also a constant, then the time complexity of LSMA can be given by *O(cn)*.

### 3.6. Algorithm GALS

Based on the discussion in above sections, the description of algorithm GALS is given as Fig. 5.

```
Procedure  GALS
Input: N, L, μ, λ  /* N denotes the complex network, L denotes the iteration number of GALS,
       μ denotes the size of parent population, λ denotes the size of offspring population */
Output: C /* network community structure, or called network clustering solution */
Begin
1    P← produce initial population by running MRW for μ times
2    For i=1: L
3        P^(new) ← ∅
4        For j=1: λ
5            g← apply uniform crossover on two arbitrary parents from P
6            g← apply LSMA mutation on g // the mutation rate is about β
7            P^(new) ← P^(new) ∪ {g}
8        End
9        P^(u) ← P ∪ P^(new)
10       P← single out μ best individuals from P^(u)  // μ+λ selection strategy
11   End
12   I← select the fittest individual from P
13   C← decode chromosome I
End
```

**Fig. 5.** GALS algorithm.

Algorithm GALS starts by adopting a Markov random walk based individual generation method (MRW) to produce the initial population. Then, it detects the network community structure by iteratively executing the following three genetic operators: uniform crossover; local search based mutation; and *μ+λ* selection. It's obvious that: the individuals in initial population produced by MRW are not only safe but also accurate and diverse; our crossover operator can effortlessly achieve its global search function by merging and/or splitting communities for LAR chromosome; our mutation operator can effectively achieve its local search function by cleverly and directly moving single marginal nodes between communities; our selection operator makes the best individuals enter the next generation, embodying Darwin's "survival of the fittest" paradigm, although it doesn't deal with single LAR chromosomes.

Moreover, the population produced by MRW is a safe one. Meanwhile, the two genetic operators (uniform crossover and local search based mutation), that can change the structure of LAR chromosome, will also not violate the safety of the safe population. Thus it will make algorithm GALS able to detect community structure in the range of a safe solution space, which is much smaller than the whole solution space. From this perspective, the search efficiency and convergence speed of algorithm GALS can both be improved.

In network $N$, let the total number of nodes be $n$, and the average community size in all of the LAR chromosomes during the execution process of algorithm GALS be $c$. It's obvious that $c$ is much smaller than $n$. The time complexity of GALS is given as follows.

**Proposition 4.** *The time complexity of algorithm GALS is $O(cn)$.*

**Proof.** It's obvious that the time complexity of the 4$^{th}$ step is the highest in GALS, and the time of other steps are all equal to or less than $O(n)$. The run time of algorithm LSMA in step 4 will be $L\lambda$, and the run time of a single LSMA run is $O(cn)$, known from Proposition 3. Thus, the run time for step 4 in GALS will be $O(L\lambda cn)$. Because each of the parameters in GALS can be regarded as a constant, the run time for step 4 in GALS can be also given by $O(cn)$. Therefore, the time complexity of GALS is $O(cn)$. □

## 4. Experiments and Evaluation

In order to evaluate the performance of algorithm GALS, we tested it on two types of benchmark artificial networks, as well as in some widely used real-world networks.

In the experiment, our GALS is compared with six representative community mining algorithms, in which FN [14], FUA [16], TGA (Tasgin's Genetic Algorithm) [23] and GACD (Genetic Algorithm for Community Detection) [28] are modularity optimization methods, while FEC [20] and LPA [17] are heuristic methods. Note that FUA has been regarded as one of the most effective community mining methods by the famous survey of Fortunato [6], and TGA and GACD both are known and current genetic algorithms for community detection.

In algorithm GALS, there are three parameters: iteration number $L$, parent population size $\mu$ and offspring population size $\lambda$, which are all standard parameters in genetic algorithm. They can be set as: $L = 500$, $\mu = 80$ and $\lambda = 60$ based on [12, 26, 27] as well as in our own experience.

All experiments are done on a single Dell Server (Intel(R) Xeon(R) CPU 5130 @ 2.00GHz 2.00GHz processor with 4Gbytes of main memory), and the source code of the algorithms used here can all be obtainable from the authors.

### 4.1. *Computer-generated networks*

We adopt two types of randomly generated synthetic networks (by both Newman model [3] and LFR model [34]) with a known community structure to evaluate the performance of the algorithms. Moreover, here we employ a widely used accuracy measure so called Normalized Mutual Information (NMI) [4]. The NMI measure, which makes use of information theory models, is regarded as a relatively fair metric compared with the other ones [4].

#### 4.1.1. *Newman benchmark*

The first type of synthetic networks employed here is that proposed by Newman et al [3]. For this benchmark, each graph consists of $n = 128$ vertices divided into 4 groups of 32 nodes. Each vertex has on average $z_{in}$ edges connecting it to members of the same group and $z_{out}$ edges to members of other groups, with $z_{in}$ and $z_{out}$ chosen such that the total expected degree $z_{in}+z_{out} = 16$, in this case. As $z_{out}$ is increased from the small initial values, the resulting graphs pose greater and greater challenges to the community mining algorithms. In Fig. 6(a), we show the NMI accuracy attained by each algorithm as a function of $z_{out}$. As we can see, our algorithm GALS outperforms all the other six methods in terms of NMI accuracy on this benchmark.

Computing speed is another very important criterion to evaluate the performance of an algorithm. Time complexity analysis for GALS has been given by proposition 4 in Sec. 3.6. Nevertheless, here we show the actual running time of GALS from an experimental angle, so as to further evaluate its efficiency.

Here we also adopt the synthetic networks based on Newman model [3]. For this application, each graph consists of $n = 100a$ vertices divided into $a$ groups of 100 nodes. Each vertex has on average $z_{in} = 10$ edges connecting it to members of the same group and $z_{out} = 6$ edges to members of other groups. The only difference between the networks used here and the former ones is that, now $z_{out}$ is fixed while the community number $a$ is changeable. Fig. 6(b) shows the trend that the running time of GALS exhibits with the network scale. As we can see, the running time of GALS is proportional to the scale of the network under the condition that the average community size of the actual community structure is about a constant. Therefore, the experiment can not only validate the correctness of Proposition 4 (the time complexity of GALS is $O(cn)$), but also shows that the average community size $c$ of all the LAR chromosomes during the running process of GALS is proportional to the average community size in the actual community structure of the network. Also, it is noteworthy that, the sizes of well-defined communities in large-scale real networks are generally much smaller than the scales of networks [35].

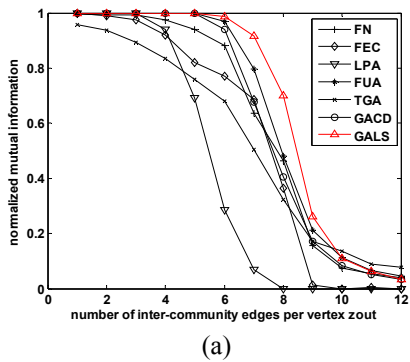

(a)

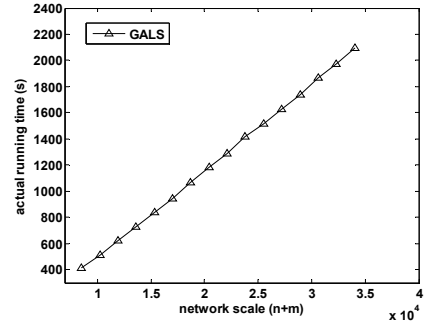

(b)

**Fig. 6.** (Color online) Test the performance of GALS on Newman benchmark. (a) Compare GALS with GN, FN, FEC, FUA, TGA and GACD in terms of NMI accuracy; (b) The actual running time of GALS is as a function of the network scale.

#### 4.1.2. *LFR benchmark*

In order to further evaluate the accuracy of these algorithms, a new type of benchmark proposed by Lancichinetti et al. [34] is also adopted here. Unlike the Newman benchmark where all the vertices have an identical degree and all the community sizes are the same, both the degree and the community size distributions in the LFR benchmark are power law, which is a statistical property that most real-world networks seem to share.

Following the experiment designed by [34], the parameters setting for the LFR benchmark networks are as follows. The network size $n$ is set to either 1000 or 5000, the minimum community size $c_{min}$ is set to either 10 or 20, and the mixing parameter $\mu$ (each vertex shares a fraction $\mu$ of its edges with vertices in other communities) varies from 0 to 0.8 with interval 0.05. We keep the remaining parameters fixed: the average degree $d$ is 20, the maximum degree $d_{max}$ is $2.5 \times d$, the maximum community size $c_{max}$ is $5 \times c_{min}$, and the exponents of the power-law distribution of vertex degrees $\tau_1$ and community sizes $\tau_2$ are -2 and -1 respectively. In Fig. 7, we show that the NMI accuracy attained by each algorithm is as a function of the mixing parameter $\mu$. As we can see, GALS is competitive with FUA, and outperforms the other five methods in terms of NMI accuracy on this new benchmark.

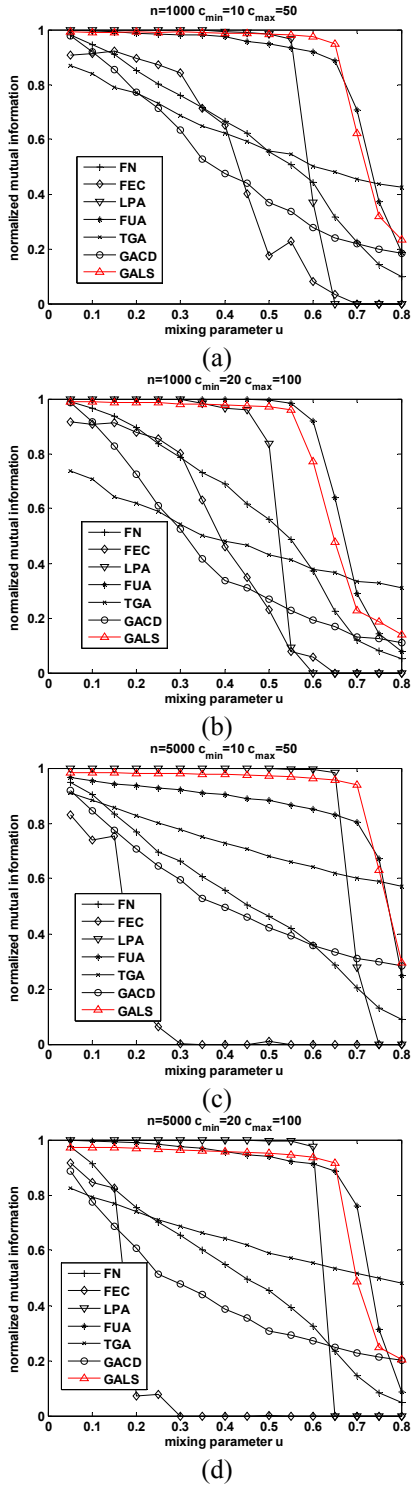

**Fig. 7.** (Color online) Compare GALS with FN, FEC, LPA, FUA, TGA and GACD in terms of NMI accuracy on the LFR benchmark networks. (a) Comparison on small networks with small communities ($n = 1000$, $c_{min} = 10$, $c_{max} = 50$). (b) Comparison on small networks with big communities ($n = 1000$, $c_{min} = 20$, $c_{max} = 100$). (c) Comparison on big networks with small communities ($n = 5000$, $c_{min} = 10$, $c_{max} = 50$). (d) Comparison on big networks with big communities ($n = 5000$, $c_{min} = 20$, $c_{max} = 100$).

### 4.2. *Real-world networks*

As real networks may have some different topological properties from the synthetic ones, here we adopt seven widely used real-world networks to further evaluate the performance of these algorithms. The networks not only include small graphs containing dozens of nodes, but also include large graphs containing tens of thousands of nodes. A description of them is given as Table 1.

Table 1. Real-world networks used in the evaluation.

| Networks | $V(N)$ | $E(N)$ | Description |
|---|---|---|---|
| karate | 34 | 78 | Zachary's karate club [36] |
| dolphin | 62 | 160 | Dolphin social network [37] |
| polbooks | 105 | 441 | Books about US politics [38] |
| football | 115 | 613 | American College football [3] |
| jazz | 198 | 5,484 | Jazz musicians network [39] |
| email | 1,133 | 5,451 | Emails of human interactions [40] |
| internet | 22,963 | 48,436 | A snapshot of the Internet [41] |

Because the inherent community structure for real networks is usually unknown, here we adopt the most commonly used modularity $Q$ [5] to evaluate the performance of these algorithms. Table 2 shows the average result (over 50 runs) that compares our method GALS with FN, FEC, LPA, FUA, TGA and GACD in terms of function $Q$ on the real-world networks described in Table 1. As we can see, the clustering quality of our method GALS is also competitive with that of FUA, and better than that of the other five algorithms.

Table 2. Compare GALS with FN, FEC, LPA, FUA, TGA and GACD in terms of $Q$ on real networks.

| $Q$-value | FN | FEC | LPA | FUA | TGA | GACD | GALS |
|---|---|---|---|---|---|---|---|
| karate | 0.3807 | 0.3744 | 0.3646 | 0.4188 | 0.4039 | 0.4198 | 0.4198 |
| dolphin | 0.5104 | 0.4976 | 0.4802 | 0.5268 | 0.5241 | 0.5294 | 0.5294 |
| polbooks | 0.5020 | 0.4904 | 0.5006 | 0.4986 | 0.5245 | 0.5272 | 0.5272 |
| football | 0.5497 | 0.5697 | 0.5865 | 0.6046 | 0.5937 | 0.6044 | 0.6046 |
| jazz | 0.4389 | 0.4440 | 0.3422 | 0.4431 | 0.4406 | 0.4435 | 0.4449 |
| email | 0.5037 | 0.5173 | 0.3706 | 0.5406 | 0.1871 | 0.4422 | 0.5599 |
| internet | 0.6378 | 0.6104 | 0.4978 | 0.6613 | 0.1141 | 0.5365 | 0.6560 |

In the seven real networks from Table 1, there are only three that have known community structures. They are the well-known karate network, dolphin network and football network. Here we try to further analyze GALS's clustering solutions in terms of the actual community structures of these three networks. Note that the running result of GALS for each of these three networks is usually the same at each time.

Zachary's karate club network [36] is a social network of friendships between 34 members of a karate club at a US university in the 1970s. Its actual community structure is shown as Fig. 8(a), in which red squares represent members associated with the principle karate teacher's faction while blue triangles represent members associated with the club administrator's faction. We randomly execute GALS once for the karate network. Its clustering solution includes four communities, which is shown as Fig. 8(a). As we can see, GALS can not only correctly discover the actual community structure of karate network, but also divide each actual community into two well-separated sub-communities. Furthermore, the average modularity $Q$ obtained by GALS over 50 runs is 0.4198, which is higher than the modularity $Q = 0.3715$ for the actual division of this network.

The dolphin network [37] describes the social relationship of 62 bottlenose dolphins living in Doubtful Sound of New Zealand, which was first established by Lusseau based on his experimental observations of the dolphins for seven years. During his research studies, he found these dolphins were separated into two groups for some reasons. The actual community structure of this network is shown as Fig. 8(b), in which the red squares represent members associated with the larger community while blue triangles represent members associated with the smaller community. We randomly execute GALS once for the dolphin network. Its clustering solution includes five communities which are shown as Fig. 8(b). As we can see, GALS can not only discover actual community structure of the dolphin network correctly, but also divide the larger actual community into four well-separated sub-communities. Furthermore, the average modularity $Q$ obtained by GALS over 50 runs is 0.5294, which is higher than the modularity $Q = 0.3722$ for the actual division of this network.

The US college football association network [3] contains 115 nodes and 613 edges, which correspond to football teams and games played among teams, respectively. All teams are divided into 12 conferences.

Each conference is considered as one actual community since the number of games played within the same conference should be much more than those between conferences. The actual community structure of this network is shown as Fig. 8(c), in which the nodes with different shapes and colors represent the football teams associated with different conferences. The names of different conferences are also given in Fig. 8(c). We randomly execute GALS once for the football network. Its clustering solution includes ten communities, which is shown as Fig. 8(c). As we can see, except for some teams in conferences Sunbelt and IA Independents, almost all other teams are correctly grouped with the other teams in their conference by our algorithm GALS. Known from [3], these misclassified teams are mostly independent teams that should not in fact belong to any conference. But, by our algorithm GALS, these independent teams are still grouped with the conference with which they are most closely associated. The few cases in which our algorithm seems to fail actually correspond to nuances in the scheduling of games. Furthermore, the average modularity $Q$ obtained by GALS over 50 runs is 0.6046, which is higher than the modularity $Q = 0.5518$ for the actual division of this network.

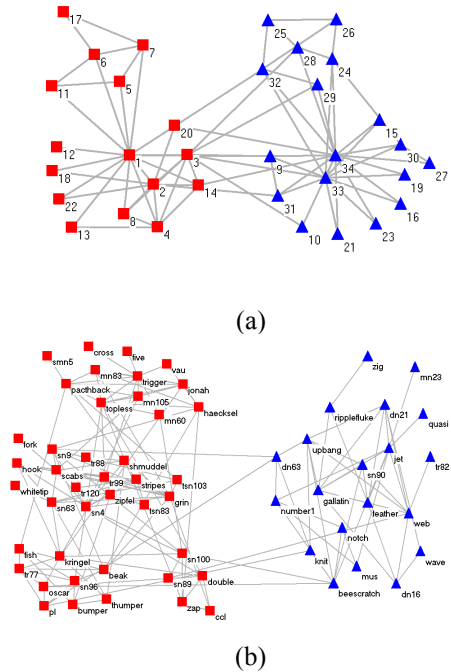

(a)

(b)

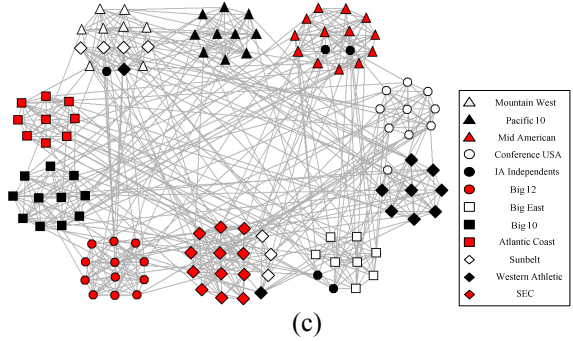

(c)

**Fig. 8.** (Color online) Clustering solutions of GALS on three widely used real-world networks with a known community structure. (a) Clustering solution of karate network; (b) Clustering solution of dolphin network; (c) Clustering solution of football network.

## 5. Conclusion

A local search based genetic algorithm (GALS), employing modularity $Q$ as objective function and LAR as genetic representation, is proposed in this paper. GALS first adopts a Markov random walk based method to produce the initial population, and then it detects community structure by iteratively executing the following three genetic operators: uniform crossover, local search based mutation and $\mu+\lambda$ selection. The proposed algorithm GALS is tested on both computer-generated and real-world networks, and compared with some presently competing community mining algorithms. Experimental results demonstrate that GALS is highly effective as well as efficient at discovering the community structure.

While GALS has a reasonable time complexity, its efficiency is still not ideal when handling huge networks (such as WWW, containing millions of nodes). This may be because the time complexity of our mutation method LSMA is $O(cn)$ which is (a little) worse than the linear time, although $c$ is much smaller than $n$. Thus, our future work can be laid as follows. We intend to improve the efficiency of LSMA from some heuristic angles, and then validate the new GALS's ability to find communities in huge networks.

Also, many real-world networks consist of communities that overlap because nodes are members of more than one community [29]. Our current work did not consider this aspect. Thus, in the future, our other task is to generalize GALS to tackle overlapping community detection, and make it able to uncover and interpret the significant overlapping in communities that is expected to be present in many cases.

**Acknowledgements**

Thanks are due to the referees for helpful comments. This work was supported by National Natural Science Foundation of China (60873149, 60973088, 61133011, 61202308), Scholarship Award for Excellent Doctoral Student granted by Ministry of Education (450060454018), Program for New Century Excellent Talents in University (NCET-11-0204), and Jilin University Innovation Project (450060481084).

**Appendix A. The proof of Theorem 1.**

**Theorem 1.** *For $\forall\, i \in V$, if the label of node i changes under the condition that the labels of all other nodes don't change, function Q of the entire network is monotonically increasing with function $f_i$.*

**Proof.** Given a network $N = (V, E)$ and its community structure $C$. For $\forall\, i \in V$, Let the label of node $i$ change from $r(i)$ to $r(j)$, which makes the community structure become $C'$. Given $r(i) \neq r(j)$, in community structure $C'$, the original community of node $i$ will become $c'_{r(i)} = c_{r(i)} - \{i\}$, and its new community will become $c'_{r(j)} = c_{r(j)} \cup \{i\}$.

From (3), we know that if any node's community changes, its function $f$ will also change. It's obvious that, the variation of node $i$'s label will result in the variation of its two related communities, which are $c_{r(i)}$ and $c_{r(j)}$. Thus, this will cause the variation of function $f$ of each node in set $c = c_{r(i)} \cup c_{r(j)}$. Here, we divide the nodes of set $c$ into three different categories, and give the equation of the variation of function $f$ for each node in each category respectively.

1) For $\forall\, s \in c'_{r(i)}$, the variation of its function $f_s$ which is defined as $\Delta f_s$ is given by

$$\Delta f_s = f_s(C') - f_s(C)$$
$$= \sum_{t \in c'_{r(i)}} \left( A_{st} - \frac{k_s k_t}{2m} \right) - \sum_{t \in c_{r(i)}} \left( A_{st} - \frac{k_s k_t}{2m} \right) \qquad \text{(A1)}$$
$$= -\left( A_{si} - \frac{k_s k_i}{2m} \right)$$

2) For $\forall\, p \in c_{r(j)}$, the variation of its function $f_p$ which is defined as $\Delta f_p$ is given by

$$\Delta f_p = f_p(C') - f_p(C)$$
$$= \sum_{q \in c'_{r(j)}} \left( A_{pq} - \frac{k_p k_q}{2m} \right) - \sum_{q \in c_{r(j)}} \left( A_{pq} - \frac{k_p k_q}{2m} \right) \quad \text{(A2)}$$
$$= A_{pi} - \frac{k_p k_i}{2m}$$

3) For node $i$, the variation of its function $f_i$ which is defined as $\Delta f_i$ is given by

$$\Delta f_i = f_i(C') - f_i(C)$$
$$= \sum_{e \in c'_{r(j)}} \left( A_{ie} - \frac{k_i k_e}{2m} \right) - \sum_{e \in c_{r(i)}} \left( A_{ie} - \frac{k_i k_e}{2m} \right) \quad \text{(A3)}$$

Thus, the variation of function $Q$ of the whole network, which is caused by the variation of the label of node $i$, can be deduced as follows. Here the variation of $Q$ is defined as $\Delta Q$.

$$\Delta Q = \frac{1}{2m} \left( \sum_{s \in c'_{r(i)}} \Delta f_s + \sum_{p \in c_{r(j)}} \Delta f_p + \Delta f_i \right) \quad \text{(A4)}$$

$$\Delta Q = \frac{1}{2m} \left( \begin{array}{l} -\sum_{s \in c'_{r(i)}} \left( A_{si} - \frac{k_s k_i}{2m} \right) \\ + \sum_{p \in c_{r(j)}} \left( A_{pi} - \frac{k_p k_i}{2m} \right) + \Delta f_i \end{array} \right) \quad \text{(A5)}$$

$$\Delta Q = \frac{1}{2m} \left( \begin{array}{l} -\left( \sum_{s \in c_{r(i)}} \left( A_{si} - \frac{k_s k_i}{2m} \right) - \left( A_{ii} - \frac{k_i k_i}{2m} \right) \right) + \\ \left( \sum_{p \in c'_{r(j)}} \left( A_{pi} - \frac{k_p k_i}{2m} \right) - \left( A_{ii} - \frac{k_i k_i}{2m} \right) \right) + \Delta f_i \end{array} \right) \quad \text{(A6)}$$

$$\Delta Q = \frac{1}{2m} \left( \begin{array}{l} \sum_{p \in c'_{r(j)}} \left( A_{pi} - \frac{k_p k_i}{2m} \right) \\ - \sum_{s \in c_{r(i)}} \left( A_{si} - \frac{k_s k_i}{2m} \right) + \Delta f_i \end{array} \right) \quad \text{(A7)}$$

$$\Delta Q = \frac{1}{2m} \left( \Delta f_i + \Delta f_i \right) = \frac{1}{m} \Delta f_i \quad \text{(A8)}$$

Then $Q(C') - Q(C) = \frac{1}{m} \left( f_i(C') - f_i(C) \right)$. As we can see, given $f_i(C') > f_i(C)$, there will be $Q(C') > Q(C)$ □